\documentclass[preprint,nofootinbib]{revtex4}

\usepackage{graphicx}
\usepackage{xcolor}
\usepackage{amsmath}

\usepackage{float} 
\usepackage{amsmath} 
\usepackage{amssymb} 
\usepackage{bbold} 

\usepackage[utf8]{inputenc} 
\usepackage[T1]{fontenc} 

\begin{document}

\title{Anisotropic models in LQC with GBP polymerisation}

\author{D. A. Cook$^1$, A. Olimpieri$^1$, I. P. R. Baranov$^2$, H. A. Borges$^1$, S. Carneiro$^{1,3}$\footnote{Corresponding author: saulocarneiro@on.br}}
\date{\today}

\affiliation{$^1$Instituto de F\'isica, Universidade Federal da Bahia, 40210-340, Salvador, BA, Brazil\\$^2$Instituto Federal de Educa\c c\~ao, Ci\^encia e Tecnologia da Bahia, 40301-015, Salvador, BA, Brazil\\$^3$Observat\'orio Nacional, 20921-400, Rio de Janeiro, RJ, Brazil}

\begin{abstract}

\vline

Polymer models are effective in describing \textcolor{black}{quantum gravity} effects around the initial singularity, leading to its replacement by bouncing surfaces on which the curvature and densities are finite. Their properties depend on the space-time symmetry and on the particular polymerisation scheme adopted. In this article we investigate anisotropic models under the Gambini-Ben\'itez-Pullin polymerisation, recently used to quantise spherically symmetric black-holes, whose interiors are isometric to Kantowski-Sachs (KS) space-times. Demanding that the minimum area defined by the bouncing surface matches the Loop Quantum Gravity area gap, we can find its radius alongside the curvature and effective density and pressures at the bounce. The density is always positive, while the pressures are negative enough to avoid the singularity. Due to the positive spatial curvature, the solution is oscillatory, reaching a maximum \textcolor{black}{radius} where a re-collapse occurs. Therefore, a positive cosmological constant is included in order to have an eternal expansion to a late de Sitter phase. We have also considered a Bianchi III metric, showing that the bounce is still present, but the space-time is asymptotically flat in this case, with no re-collapse. In this hyperbolic space, the minimal area constraint can also be imposed on compact \textcolor{black}{$2$-surfaces}. Nevertheless, in contrast to the KS case, it is enough for avoiding the singularity, independently of polymerisation procedures.

\end{abstract}

\maketitle

\section{Introduction}

Within the standard $\Lambda$-cold dark matter ($\Lambda$CDM) paradigm and classical general relativity, space-time evolution is plagued by curvature singularities: the Big-Bang in cosmology and the $r\rightarrow 0$ region inside black holes, where curvature scalars diverge. Effective polymer models, inspired by Loop Quantum Gravity (LQG) \cite{livros1,livros3}, replace these pathologies by a regular transition surface (a black-to-white hole bounce in the \textcolor{black}{latter} case), yielding non-singular space-times while recovering the classical limit far from the Planck regime \cite{Rovelli1}-\cite{worm}.

Among the various polymerisation prescriptions, a covariant one was introduced by Gambini, Benítez and Pullin (GBP) \cite{florencia} as a non-bijective canonical transformation, designed so that phase-space gauge transformations correspond to space-time diffeomorphisms. Building on this covariant spirit, Alonso-Bardaji, Brizuela and Vera (ABBV) \cite{espanhois,bascos2} constructed an effective spherically symmetric black-hole model with an asymptotically flat exterior. This latter feature solves a problem found in previous polymerisation schemes, for example the model by Ashtekar, Olmedo and Singh (AOS) \cite{PRL,AOS,guillermo2,guillermo1}, that has been shown to suffer from asymptotic non-flatness under generic conditions \cite{mariam}. Furthermore, imposing a Modesto-type minimum area condition \cite{modesto} on the transition 2-sphere ties the quantum polymer parameter to the Komar mass. Combined with semiclassical evaporation, this relation halts the black-hole mass decay at the Planck scale, yielding long-lived remnants \cite{Fernando,Fernando2,Felipe}.

In this work we apply the Loop Quantum Cosmology (LQC) techniques \cite{bojwald,LQC} to a black-hole interior, which is isometric to a homogeneous\textcolor{black}{,} Kantowski–Sachs (KS) space-time. We adopt the GBP \textcolor{black}{polymerisation} and work in the gauge \textcolor{black}{with lapse function}
$N=1$, which provides a natural cosmological time for the dynamics. Within this setting, the effective evolution becomes non-singular: the classical big-bang singularity is replaced by a bounce, and the post-bounce dynamics exhibits an oscillatory behaviour. 
It will be shown that adding a cosmological constant avoids the re-collapse, resulting in an asymptotic de Sitter universe.

For completeness, we also consider a Bianchi type III metric, the negatively-curved version of the KS metric. In this case there is a bounce, but no re-collapse, with the space-time evolving asymptotically to a Minkowski one in the absence of a cosmological constant. The bounce emerges from the imposition of Modesto's minimal area constraint on compact \textcolor{black}{$2$-surfaces}. In this case, however, it does not depend on the polymerisation procedure, appearing even in the classical version of the Hamilton equations, provided that the LQG minimal area is postulated.

The paper is organised as follows. In the next section we present the classical and polymerised Hamiltonians, deriving the Hamilton equations from the latter. In section III the solution for the metric is found and the minimal area condition is imposed on the KS scale factors. Section IV analyses the bounce and the re-collapse, showing the effective energy content around these times. The effect of introducing a positive cosmological constant is also discussed. In section V we consider the Bianchi III space-time, studying in particular the bounce and the asymptotic limit. In the Conclusions \textcolor{black}{section} we present our final remarks.

\section{The model}

In order to get a natural cosmological time, we set the lapse function $N=1$. In this way, the homogeneous metric can be written as\footnote{In the case of a black-hole interior, the lapse is usually chosen as $N = \gamma \sqrt{p_c}/b$ \cite{AOS,Fernando2}.}
\begin{equation}
    ds^{2}=-dt^{2}+\frac{p_{b}^{2}}{p_{c}}dx^{2}+p_{c}d\Omega^{2},
    \label{eq: métrica homogênea clássica}
\end{equation}
where $t$ is the cosmological time and
\begin{equation}
    d\Omega^2 = d\theta^2 + \sin^2{\theta} d\phi^2.
\end{equation}
The corresponding classical Hamiltonian is given by \cite{AOS}
\begin{equation}
    H_{\text{cl}}=\frac{1}{G}\left[-\frac{p_{b}}{2\sqrt{p_{c}}}\left(1+\frac{b^{2}}{\gamma^{2}}\right)-\frac{\sqrt{p_{c}}}{\gamma^{2}}bc\right],
    \label{eq: hamiltoniana clássica}
\end{equation}
where  $\gamma$ is the Barbero-Immirzi parameter, and the canonical variables obey the algebra $\{b,p_{b}\}=G\gamma$ and $\{c,p_{c}\}=2G\gamma$.

\textcolor{black}{This algebra is respected by} the GBP polymerisation \cite{florencia}, defined from the \textcolor{black}{canonical transformations}\footnote{\textcolor{black}{Note that the ($c$, $p_c$) sector is not polymerised. In the black-hole context, this assures that the horizon area is the same as in the classical solution \cite{Fernando,Fernando2}.}}
\begin{equation}
    b \rightarrow \frac{\sin(\delta_{b}b)}{\delta_{b}}, \quad \quad p_{b} \rightarrow \frac{p_{b}}{\cos(\delta_{b}b)}.
    \label{eq: esquema de polimerização trigonométrica}
\end{equation}
Multiplying \eqref{eq: hamiltoniana clássica} by the regularisation factor $\cos(\delta_{b}b)/b_{0}$, it leads to the effective Hamiltonian
\begin{equation}
    {H_{\text{eff}}=-\frac{1}{2G\gamma b_{0}}\left[\frac{\gamma p_{b}}{\sqrt{p_{c}}}\left(1+\frac{\sin^{2}(\delta_{b}b)}{\gamma^{2}\delta_{b}^{2}}\right)+ \frac{c\sqrt{p_{c}}\sin(2\delta_{b}b)}{\gamma\delta_{b}}\right],}
    \label{eq: hamiltoniana efetiva}
\end{equation}
where $b_{0}\equiv\sqrt{1+\gamma^{2}\delta_{b}^{2}}$.

With the above polymerisation the metric assumes the form
\begin{equation}
    {ds^{2}=-dt^{2}+\frac{p_{b}^{2}}{\cos^{2}(\delta_{b}b)p_{c}}dx^{2}+p_{c}d\Omega^{2}.}
    \label{eq: elemento de linha polimerizado}
\end{equation}
The Hamilton equations are derived as usually, leading to
\begin{align}
    \dot{b}=\{b,H_{\text{eff}}\}&=G\gamma \frac{\partial H_\text{{eff}}}{\partial p_{b}} = -\frac{\gamma}{2b_{0}\sqrt{p_{c}}}\left(1+\frac{\sin^{2}(\delta_{b}b)}{\gamma^{2}\delta_{b}^{2}}\right), \label{eq: equação de b ponto}\\
    \dot{c}=\{c,H_{\text{eff}}\}&=2G\gamma \frac{\partial H_{\text{eff}}}{\partial p_{c}}=-\frac{1}{2b_{0}\sqrt{p_{c}}}\left[-\frac{\gamma p_{b}}{p_{c}}\left(1+\frac{\sin^{2}(\delta_{b}b)}{\gamma^{2}\delta_{b}^{2}}\right)+\frac{c\,\sin(2\delta_{b}b)}{\gamma\delta_{b}}\right],\label{eq: equação de c ponto}\\
    \dot{p}_{b}= \{p_{b},H_\text{{eff}}\}&=-G\gamma \frac{\partial H_\text{{eff}}}{\partial b}=\frac{1}{2b_{0}\gamma}\left[\frac{ p_{b}}{\sqrt{p_{c}}}\frac{\sin(2\delta_{b}b)}{\delta_{b}}+ 2c\sqrt{p_{c}}\cos(2\delta_{b}b)\right],\label{eq: equação de pb ponto} \\
    \dot{p_{c}}= \{p_{c},H_\text{{eff}}\}&=-2G\gamma \frac{\partial H_\text{{eff}}}{\partial c} = \frac{1}{b_{0}}\frac{\sqrt{p_{c}}\sin(2\delta_{b}b)}{\gamma \delta_{b}},\label{eq: equação de pc ponto}
\end{align}
where the dot means derivative with respect to $t$.

\section{Solutions}

Let us obtain the solutions needed to construct the metric \eqref{eq: elemento de linha polimerizado}. First, from \eqref{eq: equação de b ponto} and \eqref{eq: equação de pc ponto} we obtain the first integral
\begin{equation}
    {M =\frac{\sqrt{p_{c}}}{2}\left(1+\frac{\sin^{2}(\delta_{b}b)}{\gamma^{2}\delta_{b}^{2}}\right),}
    \label{eq: constante de movimento de b ponto}
\end{equation}
where $M$ is a positive integration constant. \textcolor{black}{In the context of non-singular black-hole solutions, $M$ is identified with the black/white hole mass \cite{Fernando2}. In the present case, as the KS $3$-space is finite, it is natural to identify $M$ with its energy content, as will be clearer below.}

It is also possible to show that the product $p_c c$ is constant \cite{Fernando, Fernando2}, and this result does not depend on the lapse chosen. Indeed, from \eqref{eq: equação de c ponto} and \eqref{eq: equação de pc ponto} we have
\begin{equation}
\begin{split}
    (p_{c}c)\dot{}=\frac{1}{2b_{0}}\left[\frac{\gamma p_{b}}{\sqrt{p_{c}}}\left(1+\frac{\sin^{2}(\delta_{b}b)}{\gamma^{2}\delta_{b}^{2}}\right)+\frac{c\sqrt{p_{c}}\sin(2\delta_{b}b)}{\gamma\delta_{b}}\right],
\end{split}   
\end{equation}
which is equal to zero due to the Hamiltonian constraint $H_{\text{eff}} \approx 0$.

Using \eqref{eq: constante de movimento de b ponto} and the Hamiltonian constraint, we can obtain the metric central term
\begin{equation} \label{centralterm}
    \frac{p_{b}^{2}}{p_{c}\cos^{2}(\delta_{b}b)}=\frac{(cp_{c})^{2}}{M^{2}\gamma^{2}}\left(\frac{2M}{\sqrt{p_{c}}}-1\right).
\end{equation}
Setting $cp_{c} = M \gamma$, the line element \eqref{eq: elemento de linha polimerizado} can then be written as
\begin{equation} \label{line}
    {ds^{2}=-dt^{2}+\left(\frac{2M}{\sqrt{p_{c}}}-1\right)dx^{2}+p_{c}d\Omega^{2}.}
\end{equation}

\subsection{\bf Solution for ${p}_{c}$}

To get a solution for $p_c$ we first write, from \eqref{eq: constante de movimento de b ponto} and with the help of trigonometric identities,
\begin{equation}
    \begin{split}
        \sin(2\delta_{b}b) &=\pm 2\gamma\delta_{b}\sqrt{\left(\frac{2M}{\sqrt{p_{c}}}-1\right)\left[1+ \gamma^{2}\delta_{b}^{2}\left(1-\frac{2M}{\sqrt{p_{c}}}\right)\right]}.
    \end{split}
\end{equation}
Substituting into \eqref{eq: equação de pc ponto} we obtain
\begin{equation}
   \begin{split}
        \dot{r}=\pm\frac{1}{b_{0}}\sqrt{\left(\frac{2M}{r}-1\right)\left[1+ \gamma^{2}\delta_{b}^{2}\left(1-\frac{2M}{r}\right)\right]},
   \end{split}
\end{equation}
where we have defined the radial coordinate $r \equiv\sqrt{p_{c}}$. Its solution is given by the transcendental equation
\begin{equation} \label{solucao}
    {(1+2\gamma^{2}\delta_{b}^{2})\arcsin\left(\frac{b_{0}^{2}}{M}r-1-2\gamma^{2}\delta^{2}_{b}\right)-\sqrt{1-\left(\frac{b_{0}^{2}}{M}r-1-2\gamma^{2}\delta_{b}^{2}\right)^{2}}=\pm\frac{b^{2}_{0}}{M} t + C,}
\end{equation}
where $C$ is an integration constant. 

\begin{figure}[t]
\centerline{\includegraphics[scale=0.35]{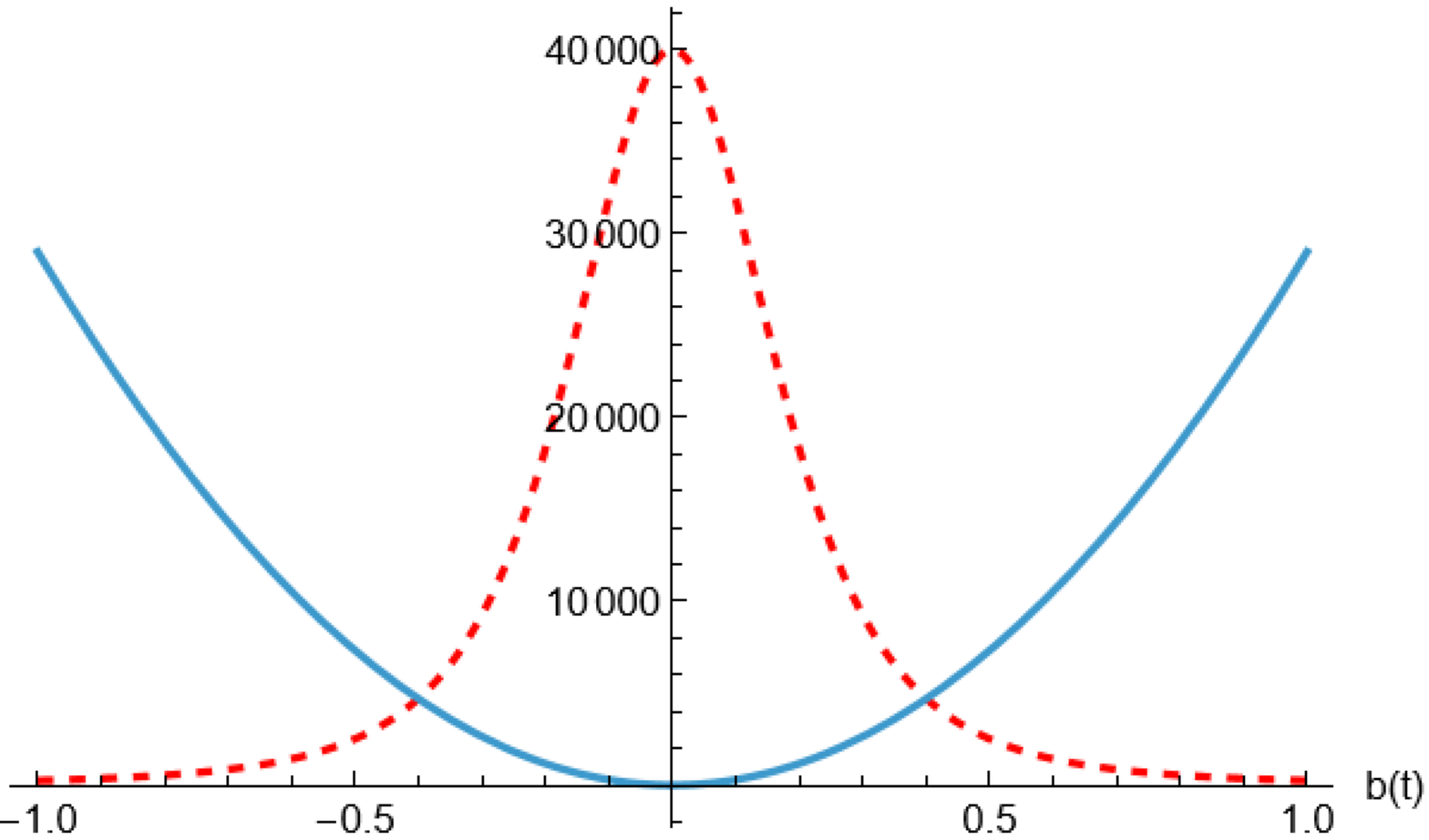}}
\caption{The squared scale factors $p_c$ (dashed red line) and $g_{xx}$ (solid blue line) of metric (\ref{line}) as functions of $b(t)$, for $M = 100$, $\gamma = \sqrt{3}/6$ and $\delta_b$ given by (\ref{area_minima}). \textcolor{black}{The constant ($cp_c$) in (\ref{centralterm}) was arbitrarily fixed.}}
\label{Fig.1}
\end{figure}

The numerical solution for $p_c$, parameterised as a function of $b(t)$, is shown in Fig.~\ref{Fig.1} for $M = 100$, $\gamma = \sqrt{3}/6$ and $\delta_b$ given by Eq.~(\ref{area_minima}) below. The figure also shows the square of the second scalar factor, $g_{xx}$ in metric (\ref{line}). \textcolor{black}{Whereas the latter goes to zero at $b=0$, $p_c$ presents
a minimum given by Eq.~(\ref{critico2}) below.}

\subsection{The minimal radius}

To further investigate what the singularity is replaced by, we can obtain the critical points for $\sqrt{p_c}$ from Eq.~\eqref{eq: equação de pc ponto}. By doing $\dot{p}_{c}=0$, it follows that $\sin(2\delta_{b}b)=0$, that is, either $\cos(2\delta_{b}b)=1$ or $\cos(2\delta_{b}b)=-1$.
With $\cos(2\delta_{b}b)=1$ in \eqref{eq: constante de movimento de b ponto}, we have
    \begin{equation}
        \sqrt{p_{c}}_{\text{crit} 1}=2M,
    \end{equation}
     while $\cos(2\delta_{b}b)=-1$ leads to
    \begin{equation} \label{critico2}
        \sqrt{p_{c}}_{\text{crit}2}=\frac{2M\gamma^{2}\delta_{b}^{2}}{b_{0}^{2}}.
    \end{equation}
As $b^{2}_{0}>\gamma^{2}\delta_{b}^{2}$, we have $\sqrt{p_{c}}_{\text{crit} 1}>\sqrt{p_{c}}_{\text{crit} 2}$. In this way, we will define $r_{\text{max}} \equiv2M$ and $r_{\text{min}} \equiv 2M\gamma^{2}\delta_{b}^{2}/b_{0}^{2}$. 

\textcolor{black}{We see that, in the KS case, the minimum radius is proportional to the polymerisation parameter, and it goes to zero when $\delta_b \rightarrow 0$. In other words, the role of quantum effects -- which are responsible for preventing the initial singularity -- is encoded in the polymerisation procedure.}
The polymerisation parameter can be fixed by imposing the Modesto's minimal area condition \cite{modesto}, which identifies the area of the \textcolor{black}{bouncing} surface with the area gap of Loop Quantum Gravity,
\begin{equation}
    4\pi r_{\text{min}}^{2} = 4\pi \sqrt{3} \gamma.
\end{equation}
Using $\gamma=\sqrt{3}/6$, we can derive\footnote{This value for the Barbero-Immirzi parameter comes from the numerical analysis of the entropy associated to microscopic horizons \cite{CQG}, but the qualitative results derived here do not depend on this choice.}
\begin{equation} \label{area_minima}
r_{\text{min}} = \frac{\sqrt{2}}{2}, \quad \quad \quad {\delta^{2}_{b}=\frac{12}{2\sqrt{2}M-1}.}
\end{equation}

\section{Kantowski-Sachs space-time}

We will consider below the limit when $M \gg 1$, for which $\delta_b \approx 0$, $b_0 \approx 1$ and $2M\gamma^2 \delta_b^2 \approx \sqrt{2}/2$. The equation of motion for $p_c$ reduces to
\begin{equation} \label{exata}
\dot{r} = \pm \sqrt{\left(\frac{2M}{r}-1\right)\left(1-\frac{r_{\text{min}}}{r}\right)}.
\end{equation}

\subsection{The bounce}

For $t \ll 1$, the solution \textcolor{black}{for $r$} can be approximated by
\begin{equation}
\sqrt{p_c} \approx \frac{\sqrt{2}}{2} + M t^2,
\end{equation}
where we have set $t = 0$ for $r = r_{\text{min}}$. In the same approximation, the line element (\ref{line}) reduces to
\begin{equation} \label{metric1}
    ds^2 = -dt^2 + \left[ \left( 2\sqrt{2}M - 1 \right) - 4 M^2 t^2 \right] dx^2 + \left( \frac{1}{2} + \sqrt{2} M t^2 \right) d\Omega^2.
\end{equation}

\subsection{The re-collapse}

Around the time of re-collapse, the solution \textcolor{black}{can now} be approximated as
\begin{equation}
    \sqrt{p_c} \approx 2M - \frac{\tilde{t}^2}{8M},
\end{equation}
were $\tilde{t} \equiv t - t_{\text{max}}$,  $t_{\text{max}} = M\pi/2$ being the re-collapse time. The metric is reduced to
\begin{equation} \label{metric2}
    ds^2 = -d\tilde{t}^2 + \left( \frac{\tilde{t}^2}{16M^2} \right) \,d{x}^2 + \left( 4M^2 - \frac{\tilde{t}^2}{2} \right) d\Omega^2.
\end{equation}

\subsection{Effective energy content}

\subsubsection{Around the bounce}

From (\ref{metric1}) we can derive the Einstein tensor and then obtain the effective energy-momentum tensor associated to the quantum space-time fluctuations. In the natural units system with $8\pi G = c = 1$, we have, for $t \ll 1 \ll M$,
\begin{eqnarray}
    \rho &=& 2 - 6\sqrt{2} M t^2,\\ 
    p_x &=& - 4\sqrt{2} M + 24 M^2 t^2,\\
    p_{\theta} &=& p_{\phi} = - \sqrt{2} M + 24 M^2 t^2.
\end{eqnarray}
From them we also have, for $t = 0$,
\begin{equation}
   \rho + p_x + 2p_{\theta} = - 6\sqrt{2} M < 0,
\end{equation}
which shows that we have a repulsion around the minimal radius, as needed to generate a bounce. 
The curvature scalar, on the other hand, has a finite maximum at the bounce, given by $R = 6 \sqrt{2} M$, 
decreasing with the expansion.

\subsubsection{Around \textcolor{black}{the} re-collapse}

Now, from (\ref{metric2}) we can obtain, for $\tilde{t} \ll 1 \ll M$,
\begin{eqnarray}
    \rho &=& p_x = \frac{\tilde{t}^2}{64M^2},\\
    p_{\theta} &=& p_{\phi} = \frac{1}{4M^2} - \frac{\tilde{t}^2}{64M^2}. \label{app. effec. pressure}
\end{eqnarray}
In this case the expansion is decelerated around the re-collapse, since we have, \textcolor{black}{at $\tilde{t} = 0$,}
\begin{equation} \label{total_energy}
    \rho + p_x + 2p_{\theta} = \frac{1}{2M^2} > 0.
\end{equation}

\subsection{The cosmological constant}

The addition of a positive cosmological constant -- or a scalar field with slow roll potential, if we want a reliable inflationary phase -- is necessary for avoiding the re-collapse and for the asymptotic isotropisation of space. It can be added to the classical Hamiltonian as \cite{KS1,KS2,Singh1,Singh2}
\begin{equation}
    H_{\text{cl}}=\frac{1}{G}\left[-\frac{p_{b}}{2\sqrt{p_{c}}}\left(1+\frac{b^{2}}{\gamma^{2}}\right)-\frac{\sqrt{p_{c}}}{\gamma^{2}}bc\right] + \frac{\Lambda}{2G}\, p_b\, \sqrt{p_c}.
    \label{eq: hamiltoniana clássica 2}
\end{equation}

If $\Lambda \ll 1$ (in Planck units), the new term is subdominant for $r \rightarrow r_{\text{min}}$ and, after polymerisation, we re-obtain the same solutions as before. In particular, Eq.~(\ref{eq: constante de movimento de b ponto}) remains valid, and hence (\ref{area_minima}) follows again after applying the minimal area condition. For $M\gg 1$, it leads to $\delta_b \ll 1$ and $b_0 \approx 1$. Therefore, for large $r$ the new Hamilton equations \textcolor{black}{can be approximated by the classical ones},
\begin{eqnarray}
\dot{b} &=& -\frac{\gamma}{2\sqrt{p_c}} \left( 1 + \frac{b^2}{\gamma^2} \right)  + \frac{\gamma \Lambda}{2} \sqrt{p_c},\\
\dot{c} &=& -\frac{1}{2\sqrt{p_{c}}}\left[-\frac{\gamma p_{b}}{p_{c}}\left(1+\frac{b^2}{\gamma^{2}}\right)+\frac{2cb}{\gamma}\right] + \frac{\gamma \Lambda}{2} \frac{p_b}{\sqrt{p_c}},\\
\dot{p}_{b} &=& \frac{1}{\gamma}\left[\frac{p_{b} b}{\sqrt{p_{c}}} + c\sqrt{p_{c}}\right],\\
\dot{p_{c}} &=& \frac{2b\sqrt{p_{c}}}{\gamma}.
\end{eqnarray}

The first integral $M$ is now given by
\begin{equation} \label{first integral with Lambda}
    M = \frac{\sqrt{p_c}}{2} \left( 1 + \frac{b^2}{\gamma^2} \right) - \frac{\Lambda}{6} p_c^{3/2}.
\end{equation}
The extremum of $p_c$ corresponds to $b=0$, which, from the above first integral, is the real root of
\begin{equation}
    \frac{\Lambda}{3} r^3 - r + 2M = 0.
\end{equation}
\textcolor{black}{The condition for having no re-collapse is that there should be no positive root for the above equation. This is always the case, provided that $\Lambda > 1/(9M^2)$.} 

It is straightforward to verify that the above set of Hamilton equations is identically satisfied, in the limit $t \rightarrow \infty$, by
\begin{equation}
    p_c = p_b = e^{2\lambda t}, \quad \quad c = b = \textcolor{black}{\gamma \lambda}\, e^{\lambda t},
\end{equation}
where $\lambda = \sqrt{\Lambda/3}$. The Hamiltonian constraint $H_{\text{eff}} \approx 0$ \textcolor{black}{and the first integral (\ref{first integral with Lambda}) are} also fulfilled in this limit, as should be.  Therefore, metric (\ref{eq: elemento de linha polimerizado}) is reduced to
\begin{equation}
    ds^2 = -dt^2 + e^{2\lambda t}\, \left( dx^2 + d\Omega^2 \right),
\end{equation}
locally isometric to a de Sitter space-time with Ricci scalar $R = 4\Lambda + 2e^{-2\lambda t} \approx 4\Lambda$.

\textcolor{black}{This is a good point to discuss the interpretation of the integration constant $M$,
by looking at its definition in the presence of a cosmological constant, Eq.~(\ref{first integral with Lambda}). In the asymptotic de Sitter limit, the second term on the right-hand side is precisely (minus) the energy stored in the cosmological constant inside a sphere of radius $r = \sqrt{p_c}$, since the energy density is given by $\Lambda/(8\pi)$ and the de Sitter volume is $4\pi r^3/3$. Therefore, from (\ref{first integral with Lambda}) $M$ can be interpreted as the energy content in the KS space when the cosmological constant is negligible.}

\section{Bianchi III space-time}

\textcolor{black}{In the Kantowski-Sachs space-time, the re-colapse results from its positive spatial curvature, which requires the addition of a cosmological constant in order to obtain an eternal expansion. This necessity can be avoided if we consider, instead, a Bianchi III space-time, the counterpart of the KS metric with negative curvature.} In this case, in metric (\ref{eq: métrica homogênea clássica}) we have
\begin{equation}
    d\Omega^2 = d\theta^2 + \sinh^2{\theta} d\phi^2,
\end{equation}
and the classical Hamiltonian is given by \cite{Singh3}
\begin{eqnarray}
    H_{\text{cl}}=\frac{1}{G}\left[-\frac{p_{b}}{2\sqrt{p_{c}}}\left(-1+\frac{b^{2}}{\gamma^{2}}\right)-\frac{\sqrt{p_{c}}}{\gamma^{2}}bc\right].
\end{eqnarray}

Following the same polymerisation procedure as in the Kantowski-Sachs case, we obtain the Hamilton equations
\begin{eqnarray}
    \dot{b} &=& -\frac{\gamma}{2b_{0}\sqrt{p_{c}}}\left(-1+\frac{\sin^{2}(\delta_{b}b)}{\gamma^{2}\delta_{b}^{2}}\right), \label{eq: equação de b ponto 2}\\
    \dot{c} &=& -\frac{1}{2b_{0}\sqrt{p_{c}}}\left[-\frac{\gamma p_{b}}{p_{c}}\left(-1+\frac{\sin^{2}(\delta_{b}b)}{\gamma^{2}\delta_{b}^{2}}\right)+\frac{c\,\sin(2\delta_{b}b)}{\gamma\delta_{b}}\right],\label{eq: equação de c ponto 2}\\
    \dot{p}_{b} &=& \frac{1}{2b_{0}\gamma}\left[\frac{ p_{b}}{\sqrt{p_{c}}}\frac{\sin(2\delta_{b}b)}{\delta_{b}}+ 2c\sqrt{p_{c}}\cos(2\delta_{b}b)\right],\label{eq: equação de pb ponto 2} \\
    \dot{p_{c}} &=& \frac{1}{b_{0}}\frac{\sqrt{p_{c}}\sin(2\delta_{b}b)}{\gamma \delta_{b}}.\label{eq: equação de pc ponto 2}
\end{eqnarray}
From (\ref{eq: equação de b ponto 2}) and (\ref{eq: equação de pc ponto 2}) we can now derive the first integral
\begin{equation}
    {M =\frac{\sqrt{p_{c}}}{2}\left(1-\frac{\sin^{2}(\delta_{b}b)}{\gamma^{2}\delta_{b}^{2}}\right).}
    \label{eq: constante de movimento de b ponto 2}
\end{equation}

Using (\ref{eq: equação de c ponto 2}) and (\ref{eq: equação de pc ponto 2}) we show again, with the help of the Hamiltonian constraint, that $cp_c$ is constant. And, from (\ref{eq: constante de movimento de b ponto 2}) and the Hamiltonian constraint, it follows that
\begin{equation}
    \frac{p_{b}^{2}}{p_{c}\cos^{2}(\delta_{b}b)}= 1-\frac{2M}{\sqrt{p_{c}}},
\end{equation}
where we have made $cp_{c} = M\gamma$. Thus, the Bianchi-III line element assumes the form
\begin{equation} \label{line'}
    {ds^{2}=-dt^{2}+\left(1 - \frac{2M}{\sqrt{p_{c}}}\right)dx^{2}+p_{c} \left( d\theta^2 + \sinh^2{\theta} d\phi^2 \right).}
\end{equation}

\subsection{Minimal radius}

From (\ref{eq: constante de movimento de b ponto 2}) and (\ref{line'}), we can see that $p_c$ is unbounded from above, while $r \equiv \sqrt{p_c}$ has a minimum value
\begin{equation}
    r_{\text{min}} = 2M.
\end{equation}
In this hyperbolic space, slices of constant $r$ have infinite area. Therefore, in order to properly apply the Modesto's minimal area condition \cite{modesto}, we construct a compact \textcolor{black}{orientable $2$-surface} of genus $g \ge 2$, \textcolor{black}{with area} \cite{topologia}
\begin{equation}
{\cal A}_{\Sigma_g} = 4\pi (g-1) p_c.
\end{equation}
Its area is minimal for $g = 2$ and, when identified with the LQG area gap $\delta {\cal A} = 4\pi \sqrt{3} \gamma$ (with $\gamma = \sqrt{3}/6$), leads again to $r_{\text{min}} = \sqrt{2}/{2}$, alongside $M = \sqrt{2}/{4}$. 

However, it is worth of note that, oppositely to the Kantowski-Sachs case, the minimal area condition does not impose any constraint on the polymerisation parameter $\delta_b$, which can even be made zero. In other words, the bounce we will find below at $r = r_{\text{min}}$ does not result from the polymerisation procedure. It results from the imposition of the LQG minimal area \textcolor{black}{constraint} to \textcolor{black}{$2$-surfaces} constructed by compactification. If we choose $M = 0$, we would have a singularity at $r = 0$, as will be clear in what follows. For $M > 0$, on the other hand, a non-singular space-time \textcolor{black}{emerges} from the classical equations of motion.

\textcolor{black}{In this way, the interpretation of the integration constant $M$ is completely different in the Bianchi III context as compared to the Kantowski-Sachs case. On one hand, as the hyperbolic space is infinite, it cannot represent a total energy content. On the other hand, the minimum radius in the Bianchi III space-time originates from the postulation of the LQG minimal area, independently of any polymerisation scheme. The quantum gravity effects are, in this case, encoded only in the LQG area quantisation and, in this sense, $M$ plays the role of an effective quantum gravity parameter, in the same way as the polymerisation parameter in the Kantowski-Sachs solution.\footnote{\textcolor{black}{Incidentally, the value $M = \sqrt{2}/4$, found when applying the minimal area condition, coincides with the lower bound value in the KS solution, obtained from (\ref{area_minima}) in the limit $\delta_b \rightarrow \infty$.}}}

\subsection{Solutions}

\begin{figure}[t]
\centerline{\includegraphics[scale=0.3]{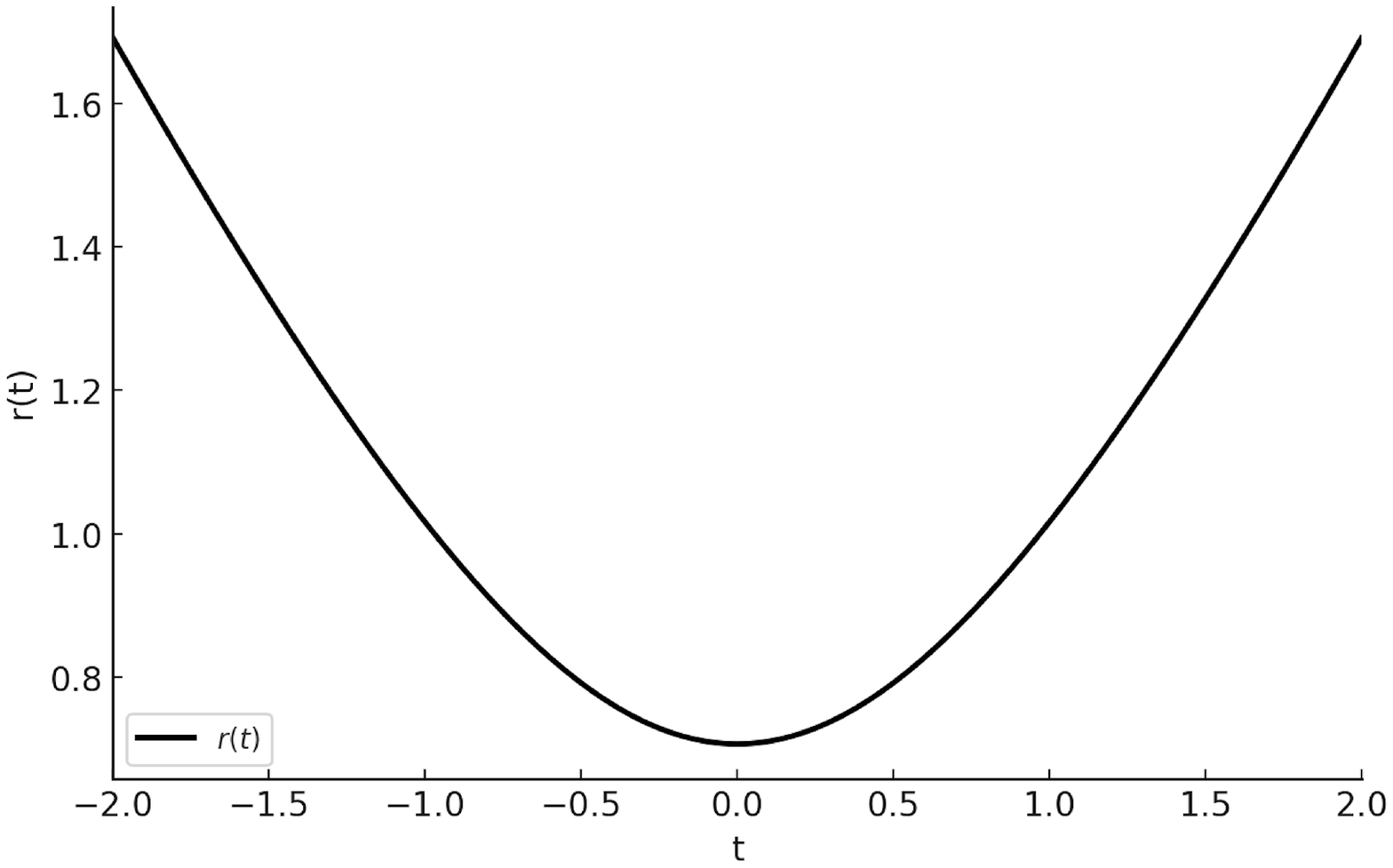}}
\caption{Time evolution of $r = \sqrt{p_c}$ for the Bianchi III solution.}
\label{Fig.2}
\end{figure}

For $\delta_b = 0$, equations (\ref{eq: equação de pc ponto 2}) and (\ref{eq: constante de movimento de b ponto 2}) lead to
\begin{equation}
    \dot{r} = \pm \sqrt{1 - \frac{2M}{r}}.
\end{equation}
Its numerical solution, for $M = \sqrt{2}/4$ and $r(0) = r_{\text{min}}$, is shown in Fig.~\ref{Fig.2}.

\subsubsection{Late times}

In this limit we have, apart an integration constant, the solution $r(t) = \pm t$, and \textcolor{black}{the} metric reduces to
\begin{equation}
    ds^2 = -dt^2 + dx^2 + t^2 \left( d\theta^2 + \sinh^2{\theta} d\phi^2 \right),
\end{equation}
\textcolor{black}{which is isometric to Minkowski space-time.}

\subsubsection{Around the bounce}

The solution is now
\begin{equation}
    r \approx 2M + \frac{t^2}{8M},
\end{equation}
and the metric can be approximated by
\begin{equation}
    ds^2 = -dt^2 + \left( \frac{t}{4M} \right)^2 dx^2 + \left( 2M + \frac{t^2}{8M} \right)^2 \left( d\theta^2 + \sinh^2{\theta} d\phi^2 \right).
\end{equation}

The effective energy density and pressures are given by
\begin{eqnarray}
    \rho &=& - p_x = \frac{8t^2}{(16M^2 + t^2)^2},\\
     p_{\theta} &=& p_{\phi} = - \frac{4}{16M^2 + t^2},
\end{eqnarray}
while for the curvature scalar we have
\begin{equation}
    R = \frac{8(16M^2 + 3t^2)}{(16M^2+t^2)^2}.
\end{equation}

We see that, for $M = 0$, there is a physical singularity at $t = 0$. On the other hand, for $M>0$ it is replaced by a bounce with finite curvature $R = 1/(2M^2)$. At the bounce,
\begin{equation}
    \rho + p_x + 2p_{\theta} = -\frac{1}{2M^2} < 0,
\end{equation}
which avoids the singularity.

\section{Conclusions}

In view of the difficulties in finding exact solutions in the realm of Loop Quantum Gravity, the effective approach through polymerisation has been shown useful for the construction of non-singular models in both contexts of black-hole and cosmological space-times. These models retain fundamental features and predictions of the full theory, in particular the existence of a minimal area and \textcolor{black}{the replacement of canonical variables by periodic functions that encode the operator nature of the holonomies.}

In the black-hole context, a proper choice of the polymerisation and quantisation schemes is essential to guarantee some desired features of the solutions, as the asymptotic flatness, the symmetry between the black and white-hole phases, that the transition surface is always inside the horizon, and that Planck mass remnants survive to the Hawking evaporation process. In this way, a promise polymerisation scheme, proposed in \cite{florencia}, was used to built suitable solutions with all these features \cite{espanhois,Fernando2}.

In these solutions the black-hole interior is isometric to Kantowski-Sachs space-times, which naturally suggests to consider the same polymerisation in a cosmological \textcolor{black}{setting}. In isotropic models of Loop Quantum Cosmology, the initial singularity is replaced by a bounce, effectively described through an additional term in the Friedmann equation. In the KS case, \textcolor{black}{we have found in the present paper that} the GBP polymerisation, combined with the identification of the bouncing surface with the LQG area gap, leads to a bounce at the Plank scale, followed by a re-collapse.

The scale of the latter depends on the integration constant $M$ characteristic of our solution. The re-collapse can be avoided by the introduction of a positive cosmological constant, provided that it is large enough as compared to $1/M^2$. In this case the metric tends asymptotically to an empty, de Sitter space. 

The bounce, on the other hand, is driven by space-time quantum fluctuations, to which we can associate an effective energy-momentum tensor with help of the Einstein equations. The energy density remains finite and positive at the bounce, satisfying in this way the weak energy condition. The pressures in turn are highly negative, giving origin to a repulsion strong enough to avoid the singularity and generate the bounce.

This qualitative behaviour is also observed when we consider a Bianchi III metric instead of the Kantowski-Sachs one. Because of its hyperbolic \textcolor{black}{geometry}, the minimal area constraint can only be imposed after the construction of appropriate compact \textcolor{black}{$2$-surfaces}. It leads again to a bounce at the Planck scale, but with no re-collapse this time, with the metric tending asymptotically to a flat space-time. Curiously enough, the bounce emerges only from the LQG minimal area constraint, not depending in this case on any polymerisation procedure.

\section*{Acknowledgements}

We are thankful to Sthefanny Rupf for following our discussions. DAC and AO thank CAPES and CNPq (Brazil) for their grants. SC is partially supported by CNPq with grant 308518/2023-3.

\end{document}